\documentclass[journal=ancac3,manuscript=article,layout=twocolumn]{achemso}
\setkeys{acs}{articletitle = true} 

\UseRawInputEncoding
\usepackage[utf8]{inputenc}
\usepackage{array}
\usepackage{stfloats}
\usepackage{comment}
\usepackage{chemformula} 
\usepackage[T1]{fontenc} 
\usepackage{float}
\usepackage{seqsplit} 
\usepackage{graphicx}
\usepackage{booktabs} 
\usepackage{multirow}
\usepackage{color, colortbl}
\usepackage[para,flushleft]{threeparttable}
\usepackage{ulem}
\usepackage[super]{nth}
\usepackage{amsmath}
\usepackage[font=scriptsize,labelfont=bf]{caption}
\usepackage[switch]{lineno}
\usepackage{breqn}
\usepackage{adjustbox}
\usepackage{makecell} 
\DeclareUnicodeCharacter{2009}{\,} 
\DeclareUnicodeCharacter{00C9}{\'E}
\DeclareUnicodeCharacter{00FC}{\"u}
\usepackage{chapterbib}
\mciteErrorOnUnknownfalse
\usepackage{amssymb}
\usepackage[version=4]{mhchem}
\usepackage{subcaption}
\usepackage{bbm}
\usepackage[shortlabels]{enumitem}
\usepackage{wrapfig}
\usepackage{amsmath}

\author{Binay P. Nayak}
\affiliation{Ames National Laboratory, and Department of Chemical and Biological Engineering, Iowa State University, Ames, Iowa 50011, United States}

\author{Wenjie Wang}
\affiliation{Division of Materials Sciences and Engineering, Ames National Laboratory, Ames, Iowa 50011, United States}

\author{Honghu Zhang}
\affiliation{NSLS-II, Brookhaven National Laboratory, Upton, New York 11973, United States}

\author{Benjamin M. Ocko}
\affiliation{NSLS-II, Brookhaven National Laboratory, Upton, New York 11973, United States}

\author{Alex Travesset}
\affiliation{Ames National Laboratory, and Department of Physics and Astronomy, Iowa State University, Ames, Iowa 50011, United States}

\author{Surya K. Mallapragada}
\email{suryakm@iastate.edu}
\affiliation{Ames National Laboratory, and Department of Chemical and Biological Engineering, Iowa State University, Ames, Iowa 50011, United States}

\author{David Vaknin}
\email{vaknin@ameslab.gov}
\affiliation{Ames National Laboratory, and Department of Physics and Astronomy, Iowa State University, Ames, Iowa 50011, United States}

\title{Electrostatically Assembled Open Square and Checkerboard Superlattices}

\begin{document}
\begin{abstract}
\vspace{-0.2cm}

Programmable assembly of nanoparticles into structures other than two-dimensional hexagonal lattices remains challenging. Assembling an open checkerboard or square lattice is harder to achieve compared to a close-packed hexagonal structure. Here, we introduce a unified, robust approach to assemble nanoparticles into a diverse family of two-dimensional superlattices at the liquid–air interface. Gold nanoparticles are grafted with pH‐responsive, water‐soluble poly(ethylene glycol) chains terminating in –COOH or \ch{-NH2} end groups, enabling control over interparticle Coloumbic interactions, while the molecular weight of grafted polymer dictates its conformation. This combined control of charges and conformation enables crystallization into checkerboard, simple–square, and body‐centered honeycomb superlattices. Furthermore, tuning the pH induces structural transitions between different lattice types. This approach opens new avenues for the fabrication of colloidal superstructures with tailored architectures.  
\end{abstract}

\section{Introduction}
\vspace{-0.2cm}
Self-assembly of nanoparticles (NPs) into a two-dimensional (2D) simple square lattice remains a challenge in materials science. Unlike the densely packed hexagonal arrangement favored by attractive forces, the `open' square motif is thermodynamically less stable and thus seldom observed in natural or synthetic systems. Its three‐dimensional (3D) analog, the simple cubic lattice, occurs only in rare elemental solids, polonium being a notable example, and theoretical studies confirm that its 2D square arrays are equally uncommon.\cite{ono2020two} Consequently, realizing this geometry with NP has proven notably elusive.

Liquid-fluid interfacial assembly is a versatile technique, capable of producing various ordered 2D NP superlattices, such as complex binary hexagonal arrays.\cite{dong2010binary,zhou2022discovery,kim2022binary}  Yet, achieving simple square lattices through these routes remains a significant hurdle. Other approaches have partially addressed this challenge, yielding square or checkerboard patterns by employing highly specific interactions or pre-designed components. For example, DNA-origami has served as a scaffold for templating NPs into checkerboard arrangements \cite{liu2016self}, and the anisotropic nature of nanocubes, combined with tailored ligand strategies, has driven their organization into similar square-shaped mesoscopic structures \cite{wang2024self}. Furthermore, engineered biomacromolecules, such as proteins and DNA tiles, have been assembled into diverse 2D arrays through programmed interactions \cite{liu2011crystalline,zhang2020assembly,yang2009structure}. Although ordered non-closed-packed NP structures are emerging, the development of broadly applicable self-assembly methods that yield simple square lattices with tunable lattice constants has yet to be achieved.

Here, we introduce a robust and generalized method for assembling NPs into checkerboard, square, and hexagonal lattices. In our approach, NPs are grafted with water-soluble polymers with distinct positive and negative charged termini. These charged polymers induce NP attractions and serve as a dynamic scaffold, directing and stabilizing their assembly into the diverse 2D binary superlattices.

\begin{figure}[!hbt]
\centering 
\includegraphics[width=0.95\linewidth]{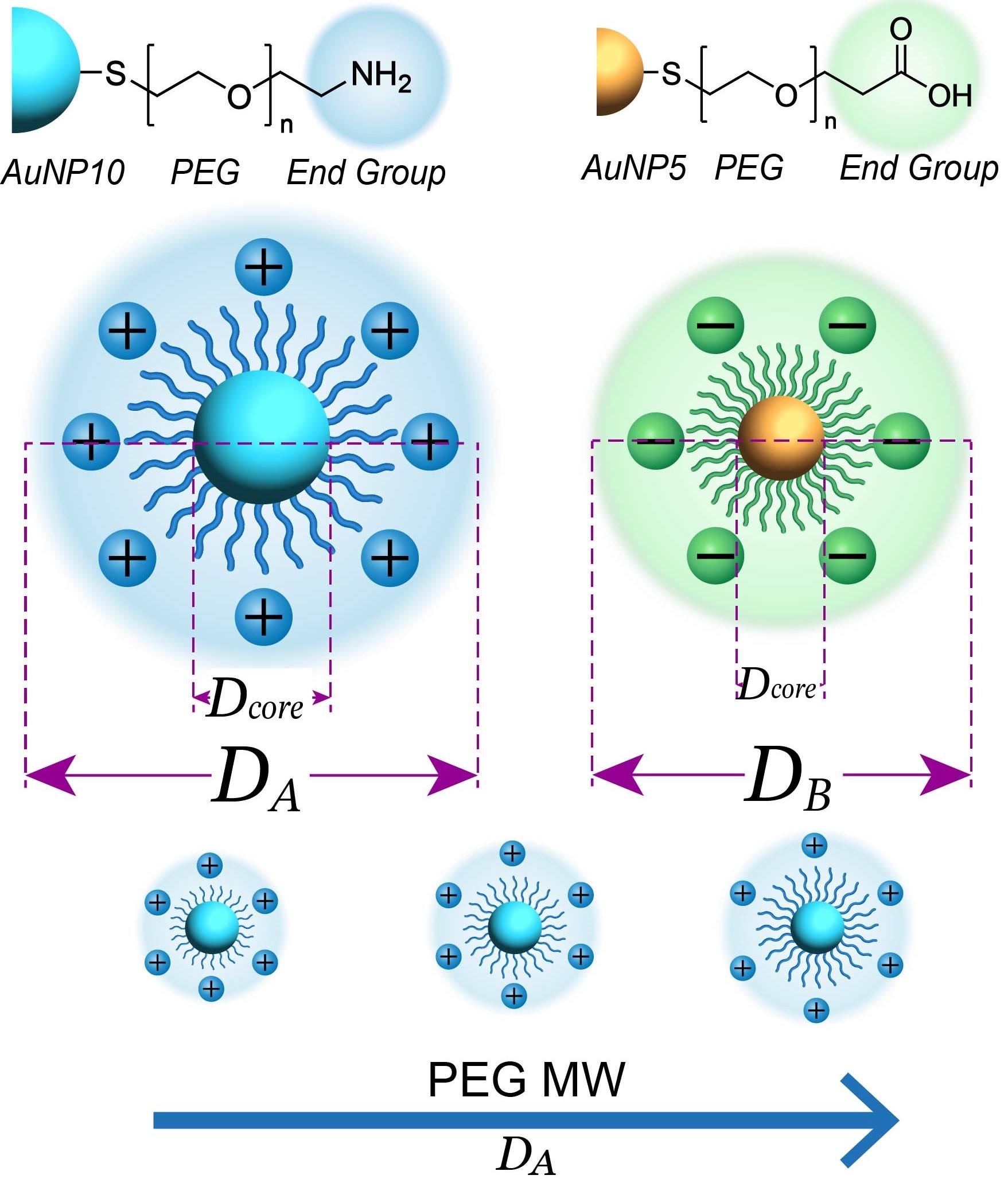}
\caption{\textbf{Schematic of Size‐ and Charge‐Tunable PEG‐Grafted AuNPs.} AuNP cores of 5 nm (yellow) and 10 nm (blue) diameter are functionalized with thiol‐PEG chains (2, 5, or 10 kDa) terminating in –NH\textsubscript{2} (blue) or –COOH (green). Positively and negatively charged PEG coronas are marked with “$+$” and “$-$” symbols. The resulting hard‐sphere diameters, \(D_A\) and \(D_B\), define the size ratio \(\gamma = D_B/D_A\). As PEG MW increases, the hydrodynamic diameter grows, as depicted in the Figure.}
\vspace{-0.4 cm}
\label{fig:scheme} 
 \end{figure}

The 2D binary superlattices presented here are directed by three tunable parameters: (i) the pH‐dependent surface charge of polymer-grafted NPs, (ii) the NP core diameter, and (iii) the molecular weight (MW) of the polymer.  The polymer MW governs the chain conformation, and thereby the hydrodynamic diameter of each particle, thus establishing an effective size ratio  
\[
\gamma = \frac{D_B}{D_A},
\]  
where \(D_A\) and \(D_B\) are the effective hydrodynamic diameters of the largest and the smallest NPs, respectively (see Fig. \ref{fig:scheme}).  This experimental strategy aligns with theoretical predictions that the NP size ratio, interaction strength, and fluid miscibility dictate 2D superlattice stability at liquid-fluid interfaces.\cite{zhou2022discovery} Extending our prior work with uniform‐MW poly(ethylene glycol) (PEG) grafted gold nanoparticles (AuNPs) that yielded honeycomb and checkerboard lattices,\cite{nayak2023ionic} we now show that the varying MW of PEG (thus $\gamma$) unlocks robust interfacial superstructures. Furthermore, modulating the surface charge via pH within the same binary system enables reversible switching between checkerboard and body‐centered honeycomb superlattices. The polymer matrix that directs these NP superlattices is optically inert, ensuring that photonic and plasmonic responses arise primarily from the spatial arrangement and intrinsic properties of the NP cores. This strategy closely parallels the three-dimensional assembly of oppositely charged nanoparticles, wherein the particles adopt an arrangement analogous to the rock-salt (NaCl-type) crystal structure. Consequently, the observed checkerboard phase, while shaped by the polymer coating, functions as a simple square template for the same core NPs, establishing the desired 2D simple square superstructure.

To create these diverse superlattices in a controlled, systematic way, we use spherical AuNPs functionalized with thiol-PEG ligands terminating in –COOH or –NH\textsubscript{2} across a range of MWs (see Fig. \ref{fig:scheme}). By adjusting the pH of the binary suspensions, we tune the surface charge of NPs that drives their organization at the air–water interface. Surface-sensitive synchrotron X-ray reflectivity and grazing‐incidence small‐angle X‐ray scattering (GISAXS) measurements determine the in-plane organization and film profile of the 2D film across the interface. Because both the NP core and the terminal ligand chemistry can be varied independently, this approach offers a broadly applicable strategy for programming 2D NP superlattices via complementary electrostatic interactions.

\vspace{-0.5cm}
\section{Results and Discussion}
\vspace{-0.2cm}

\subsubsection{From Checkerboard to Square Superlattices}
\begin{figure*}[!hbt]
\centering 
\includegraphics[width=0.9\linewidth]{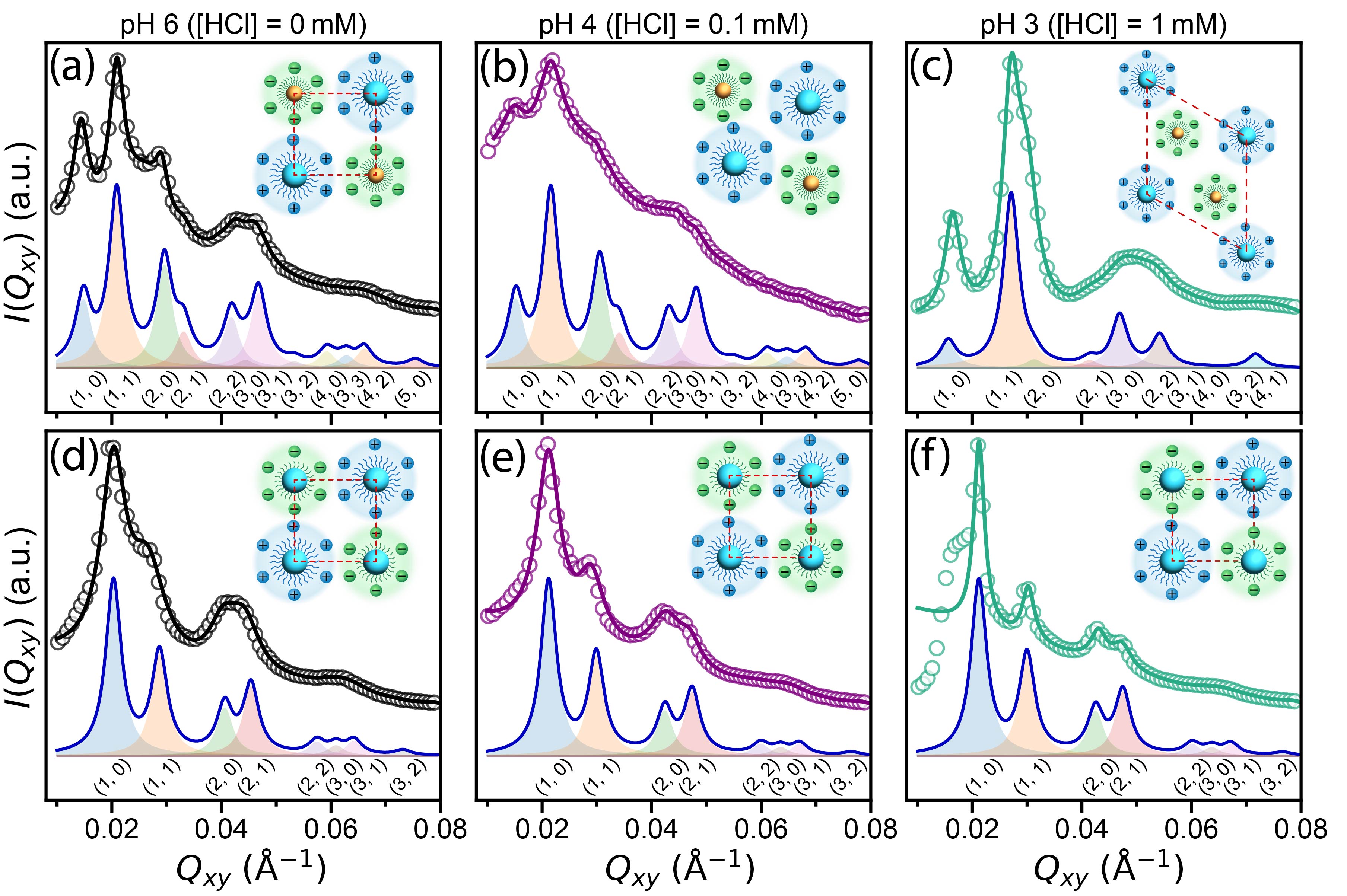}
\caption[\textbf{Checkerboard and Square Superlattices Revealed by GISAXS.}]{%
\textbf{Checkerboard and Square Superlattices Revealed by GISAXS.} Grazing‐incidence small‐angle X‐ray scattering line-cuts, $I(Q)$, for nanoparticle mixtures as a function of pH. 
(a–c) A 1:1 mixture of COOH‐PEG2k–Au5 and \ch{NH2}‐PEG5k–Au10: (a) at pH 6, a well‐defined checkerboard superlattice arises from complementary electrostatic interactions; (b) at pH 4, checkerboard order is distorted; and (c) at pH 3, a body‐centered honeycomb (quasi‐stoichiometric A$_2$B) superlattice forms. 
(d–f) Core‐matched AuNPs (10 nm) with identical polymer ligands to those in (a-c): (d) at pH 6, a perfect simple square lattice is formed; (e) at pH 4, improved square lattice ordering is induced; and (f) at pH 3, a square lattice diffraction pattern with a shoulder at lower $Q$ value indicative of partial core‐particle fractionalization. 
The solid line through the data points (open circles) is a fit to the structure factor of each lattice. Colored lines below represent the calculated structure factors, with shaded regions indicating individual peak contributions. Calculated structure factors are identical for all instances of a given structure type and are shown to illustrate the structural motifs that are intended as a qualitative guide rather than a quantitative fit to the diffraction data. The inset shows a schematic of the ideal superlattice geometry. PEGylated polymers act as scaffolds to direct nanoparticle placement into these targeted architectures.%
}
\vspace{-0.4 cm}
\label{fig:Checkerboard_Square_1} 
 \end{figure*}

\begin{figure*}[!hbt]
 	\centering 
\includegraphics[width=0.9\linewidth]{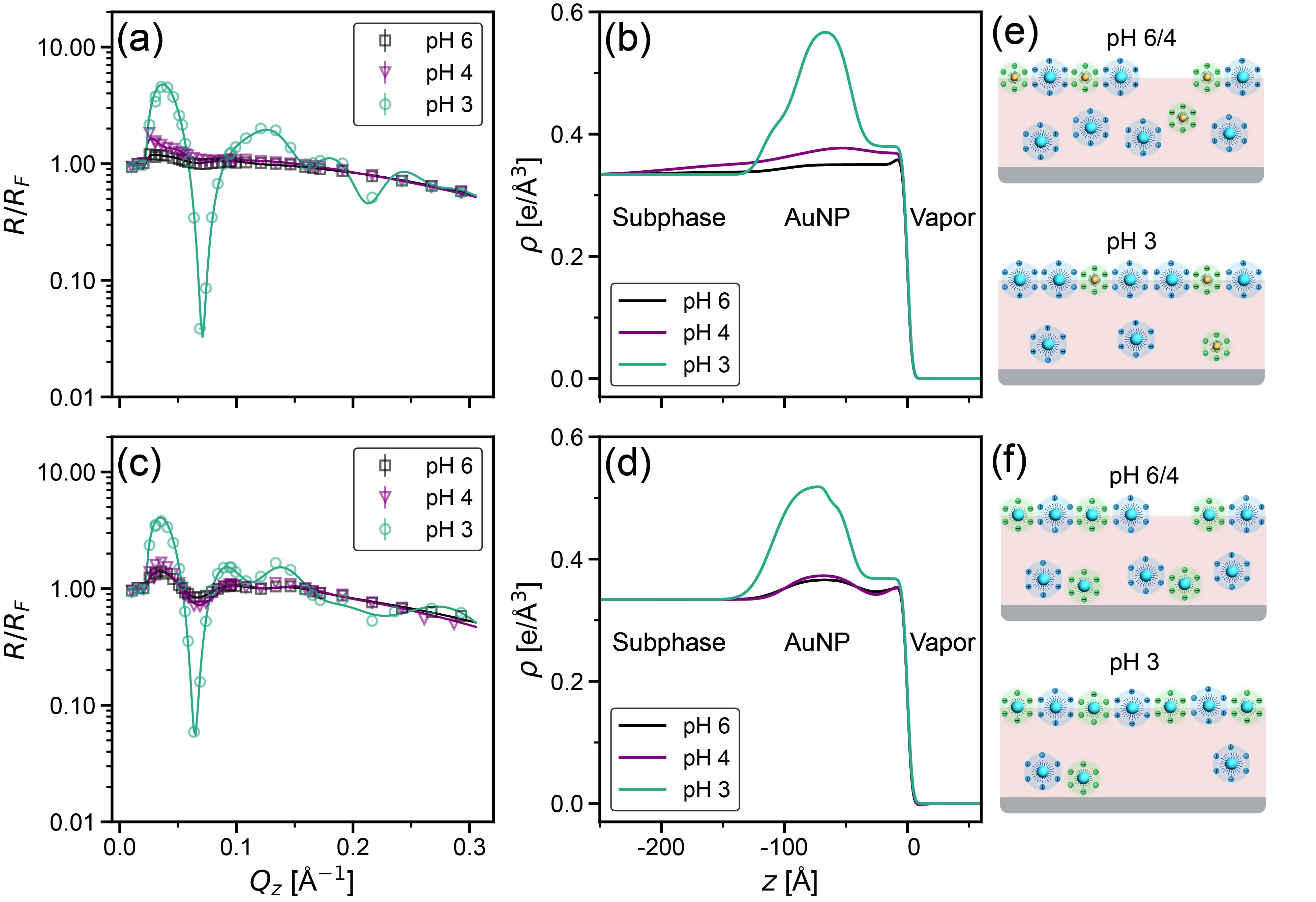}
\caption{{\bf X-ray Reflectivity Confirming Monolayer Formation and Coverage} Normalized X-ray reflectivity curves and corresponding electron density profiles, corresponding to the same samples shown in Fig. \ref{fig:Checkerboard_Square_1}.
(a) X-ray reflectivity curves for a 1:4 mixture of COOH‐PEG2k–Au5 and NH$_2$‐PEG5k–Au10 (corresponding to the samples in Fig. \ref{fig:Checkerboard_Square_1}(a-c)) at three different pH values (6, 4, and 3). Solid lines through the experimental data (symbols as indicated) represent the best fit.
(b) The electron density (ED) profiles derived from the fits in panel (a). The characteristic thickness of the excess ED layer, on the order of 10 nm, confirms that the films giving rise to the 2D diffraction patterns in Fig. \ref{fig:Checkerboard_Square_1}(a-c) consist of a single layer of AuNPs. The lower  electron density observed at pH 6 and 4 compared to pH 3 indicates incomplete surface coverage, where crystallites coexist with bare water regions, forming a `crystal-moat' morphology.
(c) X-ray reflectivity curves for the core-matched binary system of COOH‐PEG2k–Au10 and NH$_2$‐PEG5k–Au10 (corresponding to samples in Fig. \ref{fig:Checkerboard_Square_1}(d-f)), also at three pH values (6, 4, and 3).
(e-f) Schematic illustration depicting the interfacial morphology: an intact, densely packed AuNP monolayer formed at pH 3 (bottom), contrasting with dispersed crystallites coexisting with bare water `moats' at pH 4 and 6 (top).}
\vspace{-0.4 cm}
 \label{fig:Checkerboard_Square_ref} 
 \end{figure*}

To stabilize a checkerboard-like structure, we selected a binary system with a $\gamma > 0.75$, based on our earlier work with $\gamma \sim 1$.\cite{nayak2023ionic} To determine $\gamma$, we use Dynamic Light Scattering (DLS) as demonstrated in a previous publication.~\cite{nayak2025valencefree} The binary system with $\gamma = 0.77$, comprises a 1:4 mixture of COOH-PEG2k-Au5 (5 nm core) and NH$_2$-PEG5k-Au10 (10 nm core). The distinct core sizes (5 vs. 10 nm) possess sufficient X-ray diffraction electron density (ED) contrast to unequivocally identify the formation of the checkerboard lattice.
Fig. \ref{fig:Checkerboard_Square_1} (a) presents the GISAXS pattern, $I(Q_{xy}$), of the binary mixture obtained at pH 6. The well-defined diffraction pattern observed is consistent with the formation of a checkerboard superlattice. Here, the two AuNP species are grafted with PEG of distinct MW (2 kDa and 5 kDa), while our previous study used identical 5 kDa chains on both AuNPs to yield similar checkerboard lattices, albeit with distinct lattice constants.\cite{nayak2023ionic} The checkerboard lattice formed here is driven by electrostatic interactions, wherein the charges on each particle are compensated by oppositely charged neighbors, resulting in overall local charge neutrality.

For the two-dimensional checkerboard structure, the diffraction pattern typically exhibits a weak fundamental (10) peak and a stronger (11) peak. The weakness of the (10) reflection arises because its structure factor being proportional to $f_A - f_B$ ($f_A$ and $f_B$ are the form factors of two distinct AuNPs). If the two particle types are identical ($f_A = f_B$), this structure factor cancels to zero, eliminating the (10) peak. Under these conditions, the first observable diffraction peak corresponds to the (11) reflection, which then assumes the role of the fundamental (10) peak of the simple-square lattice.

Upon lowering the pH to 4 (Fig. \ref{fig:Checkerboard_Square_1} (b)), the long-range order (LRO) of the checkerboard structure deteriorates, exhibiting only short-range order (SRO) of the same structural motif. Further reduction of the pH to 3 (Fig. \ref{fig:Checkerboard_Square_1} (c)) leads to the emergence of a new diffraction pattern. The relationship between the peak positions in the GISAXS pattern indicates the formation of a hexagonal lattice with a larger unit cell than expected for a unary system. Notably, a prominent (11) peak is observed, broadly coinciding with the fundamental peak of a simple hexagonal lattice expected for individual particles. This often occurs when a new, larger hexagonal lattice originates from $\sqrt{3}\times\sqrt{3}$ expansion of the underlying single-particle arrangement.\cite{kim2022binary} 
We interpret this hexagonal lattice as a body‐centered honeycomb superlattice (quasi‐stoichiometry AB$_2$) formed by the smaller particles (5 nm core with 2 kDa PEG) arranged in a honeycomb lattice, with the larger particles (10 nm core with 5 kDa PEG) occupying the centers of the honeycomb cells (as illustrated below, and based on detailed analysis in the SI). It is worth noting that if the larger AuNPs were replaced with a low ED particle like silica, the resulting metallic framework would likely exhibit a pure honeycomb lattice, albeit with a different GISAXS diffraction pattern. Fig. S3 shows assemblies at additional mixing ratios (1:1 and 1:2), which follow the same pH‐dependent structural transitions described above. 
Transitioning from a checkerboard to a body‐centered honeycomb lattice requires increasing the molar fraction of smaller particles to achieve the 1:2 stoichiometry needed for AB\(_2\) ordering. Lowering the pH protonates the –NH\(_2\) termini on NH\(_2\)-PEG5k–Au\(_{10}\), making those particles more hydrophilic and less likely to adsorb at the interface. Under these conditions, each positively charged NH\(_2\)-AuNP is electrostatically balanced by two negatively charged COOH-AuNPs, stabilizing the AB\(_2\) honeycomb geometry.  

\begin{figure*}[!hbt]
 	\centering 
 	\includegraphics[width=0.9\linewidth]{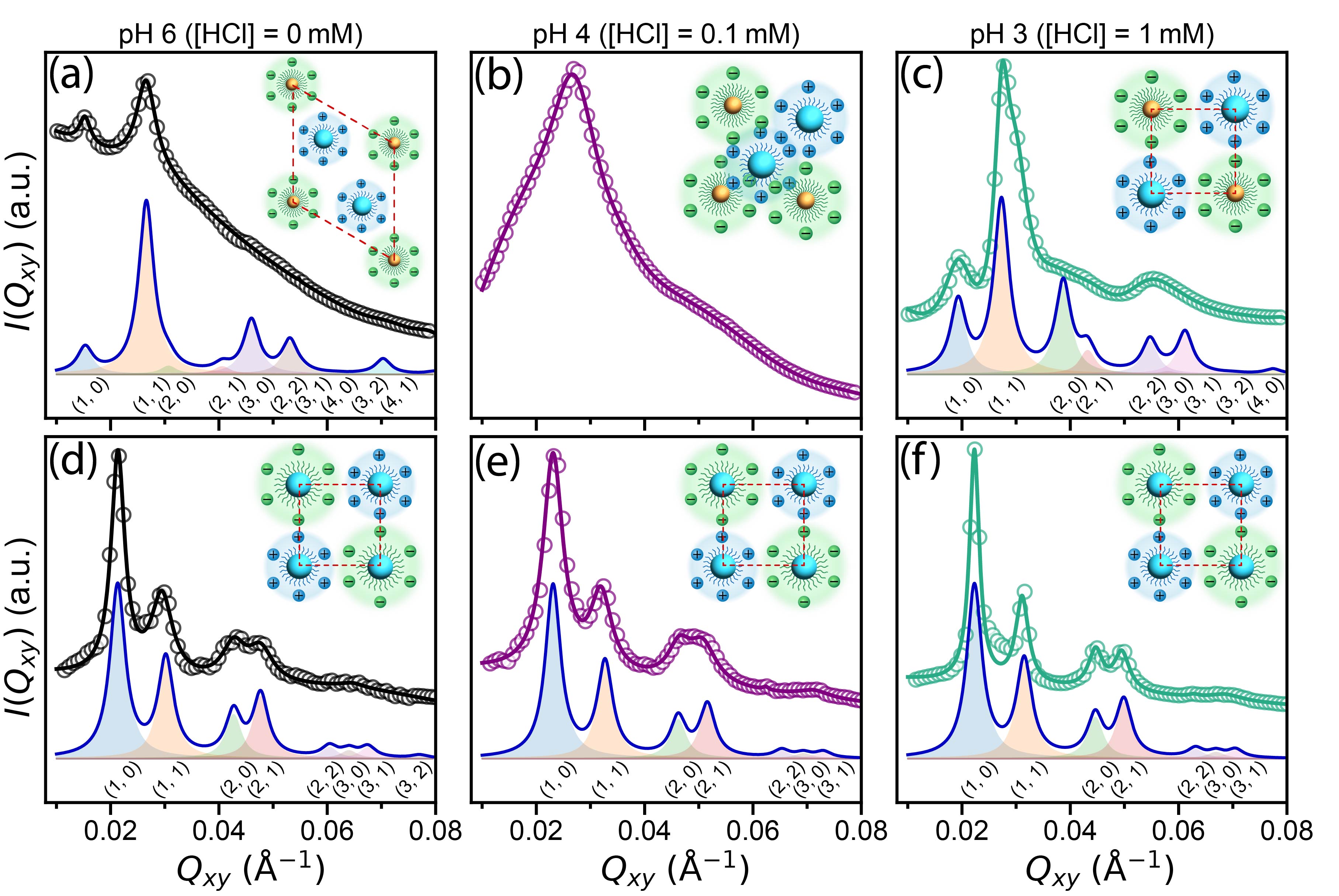}
    \caption{%
      \textbf{pH‐Dependent Lattice Transformations in PEG‐Chain‐Swapped AuNP Mixtures.} Grazing‐incidence small‐angle X‐ray scattering line-cuts, $I(Q)$, of PEG‐chain‐swapped nanoparticle systems as a function of pH. 
      (a–c) A 1:1 mixture of COOH‐PEG5k–Au5 and \ch{NH2}‐PEG2k–Au10 (PEG chains swapped relative to Fig.~\ref{fig:Checkerboard_Square_1}): (a) at pH 6, an ill‐defined body‐centered honeycomb superlattice; (b) at pH 4, short‐range ordering with two dominant Bragg reflections indicative of a disordered binary arrangement; (c) and at pH 3, a transition to an almost perfect checkerboard superlattice. 
      (d–f) Core‐matched AuNPs (10 nm) functionalized with the same swapped ligands form a robust, simple square lattice across the entire pH range: (d) at pH 6, (e) pH 4, and (f) pH 3. 
      Solid lines through the open‐circle data are fits to the corresponding structure factors; colored lines below represent the calculated structure factors, with shaded regions highlighting individual peak contributions. Calculated structure factors are identical for all instances of a given structure type and are shown to illustrate the structural motifs that are intended as a qualitative guide rather than a quantitative fit to the diffraction data. The insets illustrate the ideal superlattice geometry.%
    }
    \vspace{-0.4 cm}
 	\label{fig:Checkerboard_Square_2} 
 \end{figure*}

Establishing a robust checkerboard lattice scaffolded by the PEG matrix, we proceed to use the same configurations with the identical core size NPs (10 nm) to stabilize a simple square lattice. Fig. \ref{fig:Checkerboard_Square_1} (d–f) shows 1D GISAXS line-cuts for 1:2 binary mixtures of COOH-PEG2k-Au10 and NH$_2$-PEG5k-Au10 at pH 6, 4, and 3. At pH 6 (Fig. \ref{fig:Checkerboard_Square_1} (d)), the GISAXS pattern shows diffraction peaks that can be fitted to a simple square lattice. This square lattice ordering persists at pH 4 (Fig. \ref{fig:Checkerboard_Square_1} (e)), with an apparent enhancement in the LRO, as indicated by sharper diffraction peaks. At pH 3 (Fig. \ref{fig:Checkerboard_Square_1}(f)), a well-defined square lattice pattern is observed with even sharper and more resolved peaks. However, a faint shoulder at low \(Q\) indicates a weak ED contrast between the two nominally 10-nm populations, consistent with a square lattice exhibiting incipient checkerboard modulation. We attribute this feature to mild size fractionalization during assembly: the intrinsic size distribution within the '10 nm' particles promotes size-based sorting between sublattices, yielding subtle ED differences that manifest as the low-\(Q\) shoulder in the scattering profile.
\cite{Lai2019a,cabane2016hiding} 

To definitively determine the nature of the crystalline films observed in our GISAXS experiments, specifically whether they constitute a monolayer or multilayers, we performed complementary XRR measurements from the same films. As detailed below, and further supported by data in the Supporting Information (SI), all films discussed hereafter consist of a single NP monolayer, which can either fully or partially cover the interface. Fig. \ref{fig:Checkerboard_Square_ref} (a) presents the XRR curves obtained from a 1:4 mixture of COOH‐PEG5k-Au5 and NH$_2$‐PEG2k-Au10 at three different pH values (6, 4, and 3), corresponding to the samples analyzed by GISAXS in Fig. \ref{fig:Checkerboard_Square_1} (a-c). The solid lines through the experimental data (represented by symbols as indicated) denote the best fits as discussed in the Methods and in more detail in SI. The ED profiles derived from these fits are displayed in Fig. \ref{fig:Checkerboard_Square_ref} (b). The characteristic thickness of the excess ED layer, consistently on the order of 10 nm, provides definitive evidence that the films responsible for the 2D diffraction patterns consist of a single layer of AuNPs. Furthermore, the lower integrated excess ED observed at pH 6 and pH 4, compared to that at pH 3, indicates incomplete or lower surface coverage. This suggests a morphology where discrete crystallites coexist with bare water regions, effectively forming a 'crystal-moat' structure at these higher pH values, as illustrated schematically next to Fig. \ref{fig:Checkerboard_Square_ref}(b).

\begin{figure*}[!hbt]
 	\centering 
 	\includegraphics[width=0.9\linewidth]{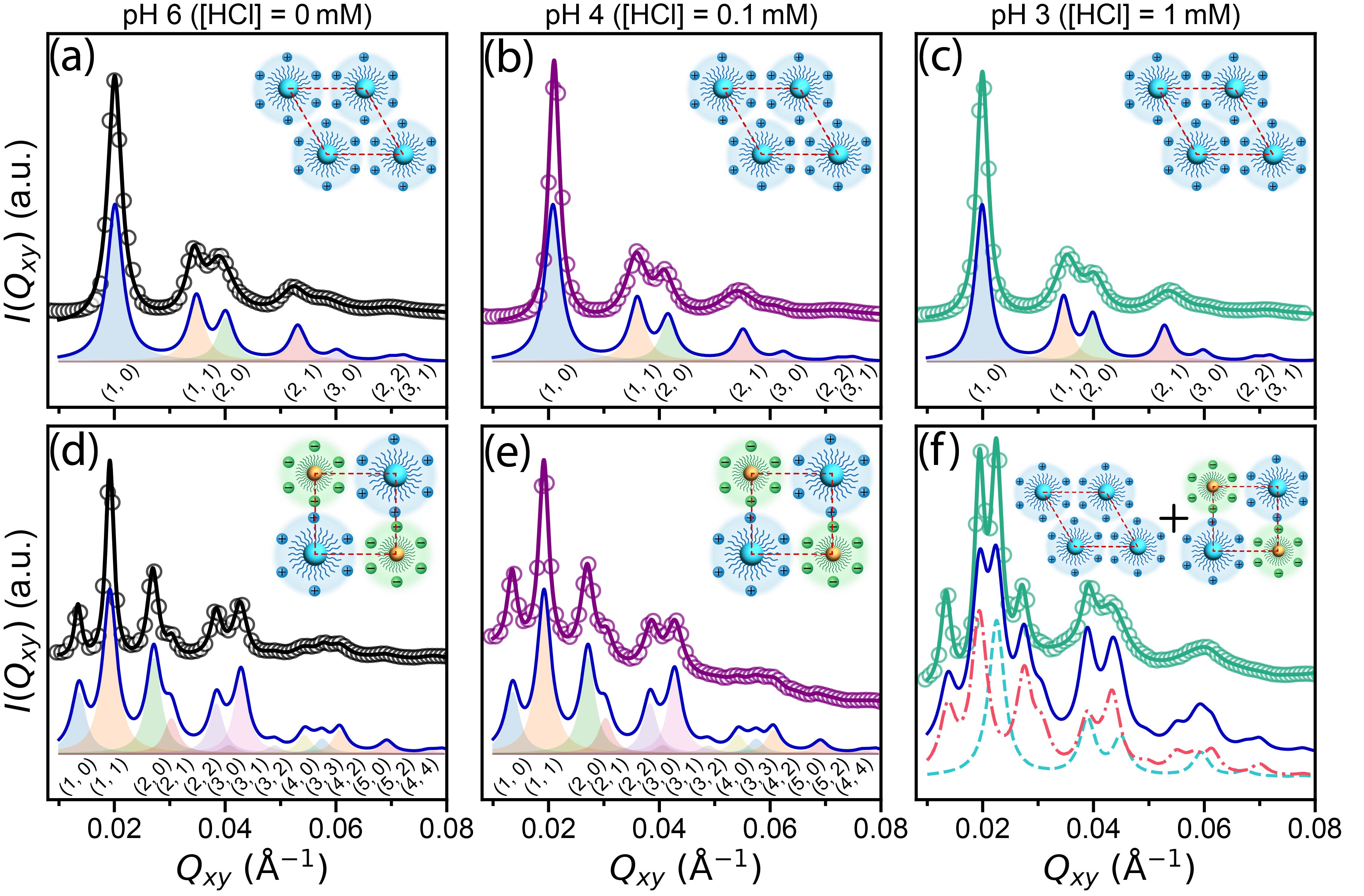}
\caption{{\bf From single-nanoparticle hexagonal lattice to checkerboard square lattice} 1D GISAXS line-cuts for a binary mixture of COOH-PEG5k-Au5 and NH$_2$-PEG10k-Au10 at three different pH values. Panels (a-c) display results for a nominal 1:1 ratio of the two components, while panels (d-f) show results for a nominal 4:1 ratio (COOH-PEG5k-Au5: NH$_2$-PEG10k-Au10).
(a-c) At a 1:1 nominal mixing ratio, the upper panels indicate that the interface is predominantly populated by a single component from the suspension, which self-assembles into a hexagonal lattice. This observation is consistent with the preferential formation of a unitary crystal by the NH$_2$-PEG10k-Au10 nanoparticles.
(d-f) As the nominal concentration of the COOH-PEG5k-Au5 is increased to a 4:1 ratio, the system adopts a checkerboard-square superlattice at both pH 6 and pH 4. However, upon lowering the pH further to 3, the interface becomes populated by a coexistence of the original unitary hexagonal superstructure and the checkerboard lattice. Solid lines through the open‐circle data are fits to the corresponding structure factors; colored lines below represent the calculated structure factors, with shaded regions highlighting individual Bragg peaks. Calculated structure factors are identical for all instances of a given structure type and are shown to illustrate the structural motifs that are intended as a qualitative guide rather than a quantitative fit to the diffraction data. The insets illustrate the ideal superlattice geometry.%
}
\vspace{-0.3 cm}
 	\label{fig:checkerboard_hexagonal_unary} 
 \end{figure*}
 
Similar XRR curves for the core-matched binary system of COOH‐PEG5k-Au10 and NH$_2$‐PEG2k-Au10 (corresponding to the samples in Fig. \ref{fig:Checkerboard_Square_1} (d-f)) are shown in Fig. \ref{fig:Checkerboard_Square_ref} (c). The derived ED profiles in Fig. \ref{fig:Checkerboard_Square_ref} (d) confirm the single-layer nature of these films as well. A schematic illustration in Fig. \ref{fig:Checkerboard_Square_ref} visually summarizes the observed interfacial morphologies: a densely packed, intact AuNP monolayer is formed at pH 3, contrasting with the morphology at pH 4 and 6, where dispersed crystallites are separated by regions of bare water, resembling 'moats'.

To probe the role of PEG MW, we swapped the polymer ligands in our binary system: COOH-PEG2k was replaced by COOH-PEG5k, and NH\textsubscript{2}-PEG5k by NH\textsubscript{2}-PEG2k, yielding COOH-PEG5k-Au5 and NH\textsubscript{2}-PEG2k-Au10. Varying the PEG chain length changes the thickness of the polymer brush, longer chains extend further into solution, while also altering grafting density. Together, these effects adjust both the spatial arrangement of the PEG (thus $\gamma$) and the magnitude of the surface charge, leading to different interfacial assembly behaviors.  
\cite{zhang2017interfacial,nayak2025effect}  

GISAXS measurements on the 1:1 COOH‐PEG5k–Au10/NH$_2$‐PEG2k–Au10 mixture ($\gamma$ = 0.98) at pH 6, 4, and 3 are shown in Fig.\ \ref{fig:Checkerboard_Square_2} (a–c). At pH 6, two distinct Bragg reflections atop a decaying background are indexed to an imperfect A$_2$B body‐centered honeycomb lattice. Lowering pH to 4 broadens and weakens these peaks, indicating loss of LRO and the emergence of only short-range correlated domains, although local honeycomb motifs remain. Further acidification to pH 3 produces multiple sharp reflections that index to a checkerboard–square superlattice. The emergence of this highly ordered state highlights the pivotal roles of polymer grafting density, tuned by PEG MWs, and the size ratio \(\gamma\) in directing assembly into specific superlattices. It is worth noting that systems with lower \(\gamma\) values stabilize a checkerboard lattice at pH 6 before transitioning to a body‐centered honeycomb lattice at pH 3, whereas higher \(\gamma\) systems exhibit the reverse sequence.  

\begin{figure*}[!hbt]
 	\centering 
 	\includegraphics[width=0.9\linewidth]{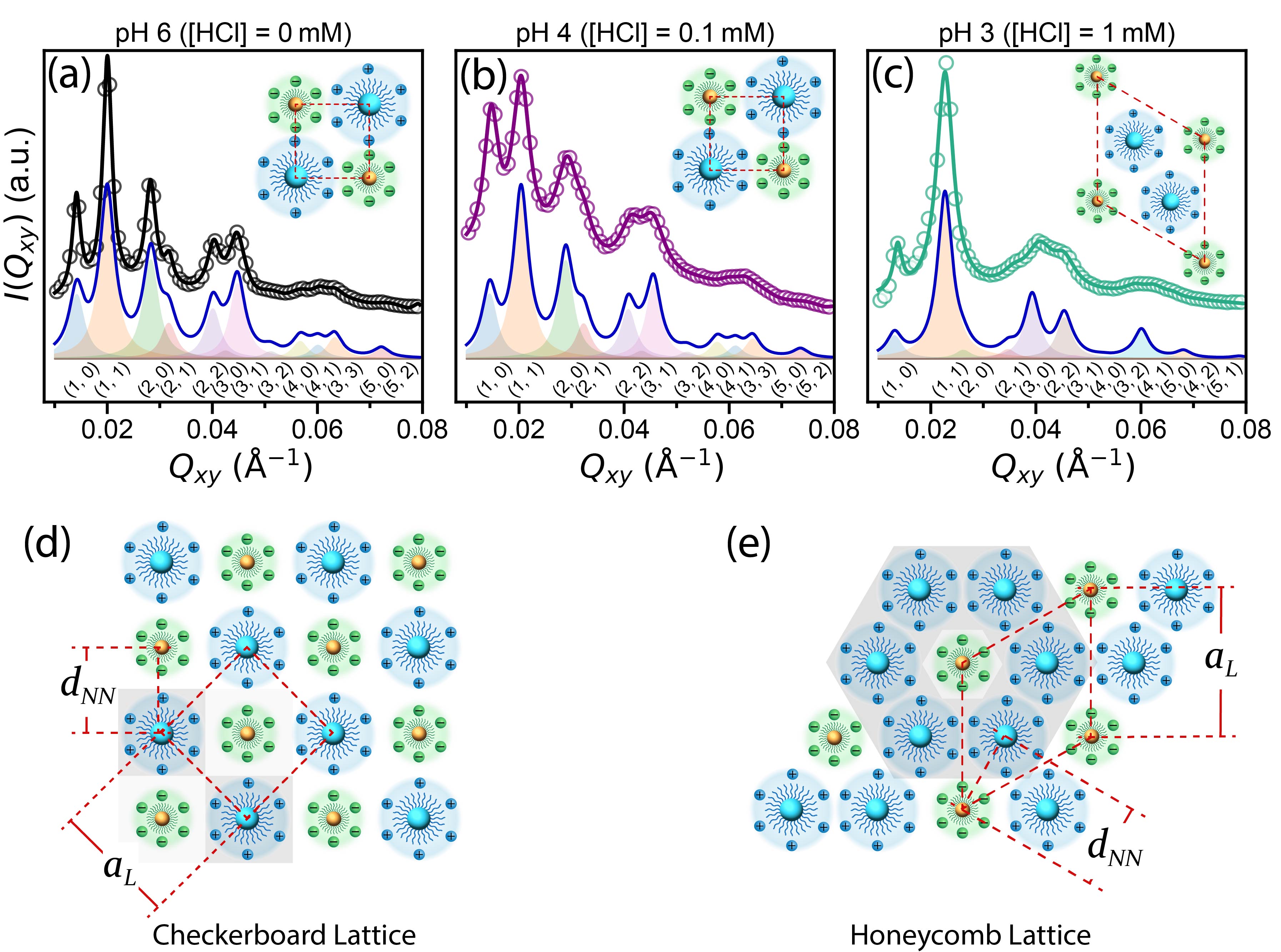}
    \caption{%
      \textbf{pH‐Driven Transition from Checkerboard to Body‐Centered Honeycomb Superlattices.}
      Grazing‐incidence small‐angle X‐ray scattering line-cuts, $I(Q)$, of a 4:1 mixture of COOH‐PEG2k-Au5 and NH$_2$‐PEG10k-Au10 as a function of pH. 
      (a) At pH 6, a well‐defined checkerboard superlattice; 
      (b) at pH 4, diminished checkerboard long‐range order; 
      (c) at pH 3, transition from checkerboard to a body‐centered honeycomb lattice with quasi‐stoichiometric A$_2$B composition. 
      Solid lines through the open‐circle data are fits to the corresponding structure factors; colored lines below represent the ideal structure factors, with shaded regions highlighting individual Bragg peaks. Calculated structure factors are identical for all instances of a given structure type and are shown to illustrate the structural motifs that are intended as a qualitative guide rather than a quantitative fit to the diffraction data. The insets illustrate the ideal superlattice geometry.
      (d–e) Top‐view schematics of (d) the checkerboard and (e) the body‐centered honeycomb lattices, with lattice constant $a_L$ and nearest‐neighbor distance $d_{NN}$ indicated.%
    }
\vspace{-0.4 cm}
 	\label{fig:Checkerboard_hexagonal} 
 \end{figure*}

To enhance the quality of the simple square lattice, we used the same PEG configuration but with identical 10 nm AuNP cores. Although different core sizes (as seen in Fig. \ref{fig:Checkerboard_Square_2} (a-c)) resulted in low-quality crystallites at the interface, using identical core particles produced perfect simple square lattices. Fig.\ \ref{fig:Checkerboard_Square_2}(d–f) presents GISAXS from a core‐matched binary mixture, both AuNPs have 10 nm cores but are grafted with COOH-PEG5k and NH$_2$-PEG2k ligands. At pH 6 (Fig.\ \ref{fig:Checkerboard_Square_2}d), the diffraction pattern corresponds to a well‐ordered simple square lattice, where the polymer serves as a scaffold for the lattice. Notably, the simple square lattice is preserved at pH 4 (Fig.\ \ref{fig:Checkerboard_Square_2}e) and exhibits even sharper diffraction peaks at pH 3 (Fig.\ \ref{fig:Checkerboard_Square_2}f), indicating improved crystallinity. As noted in the SI, the XRR from the same films is shown in the Figure with further discussion about the coverage of the films.

\subsection{pH-induced structural transitions}

Using PEG chains with a large MW mismatch (to reduce $\gamma$) not only alters the thermodynamic balance but also introduces a kinetic effect, preferential interfacial adsorption. At comparable grafting densities, AuNPs grafted with longer PEG chains (e.g., 10 kDa) experience stronger steric repulsion and a greater tendency to minimize contact with the aqueous phase than those with shorter chains (e.g., 2 kDa), leading to spontaneous interfacial population.\cite{nayak2025effect} Consequently, in a 1:1 mixture of COOH-PEG5k-Au5 and NH$_2$-PEG10k-Au10, the NH$_2$-PEG10k-Au10 particles dominate the interface, assembling into a well-ordered hexagonal lattice, as shown in Fig.~\ref{fig:checkerboard_hexagonal_unary} (a-c). This lattice persists even as the pH decreases, indicating that interfacial adsorption of the long chains is effectively irreversible on experimental timescales. In other words, the longer 10 kDa PEG corona is more hydrophobic than the shorter 2 kDa chains, thereby locking NH$_2$-PEG10k-Au10 at the interface across all pH conditions. We note that \ch{NH2}-PEG–AuNPs spontaneously assemble into hexagonal monolayers at the interface without the addition of an electrolyte, and increasing the length of the PEG chain enhances interfacial hydrophobicity, thereby promoting this early-time preferential adsorption.
\cite{kim2023two,nayak2025effect,nayak2023ionic,kim2021effect}

\begin{figure*}[!hbt]
 	\centering 
 	\includegraphics[width=0.95\linewidth]{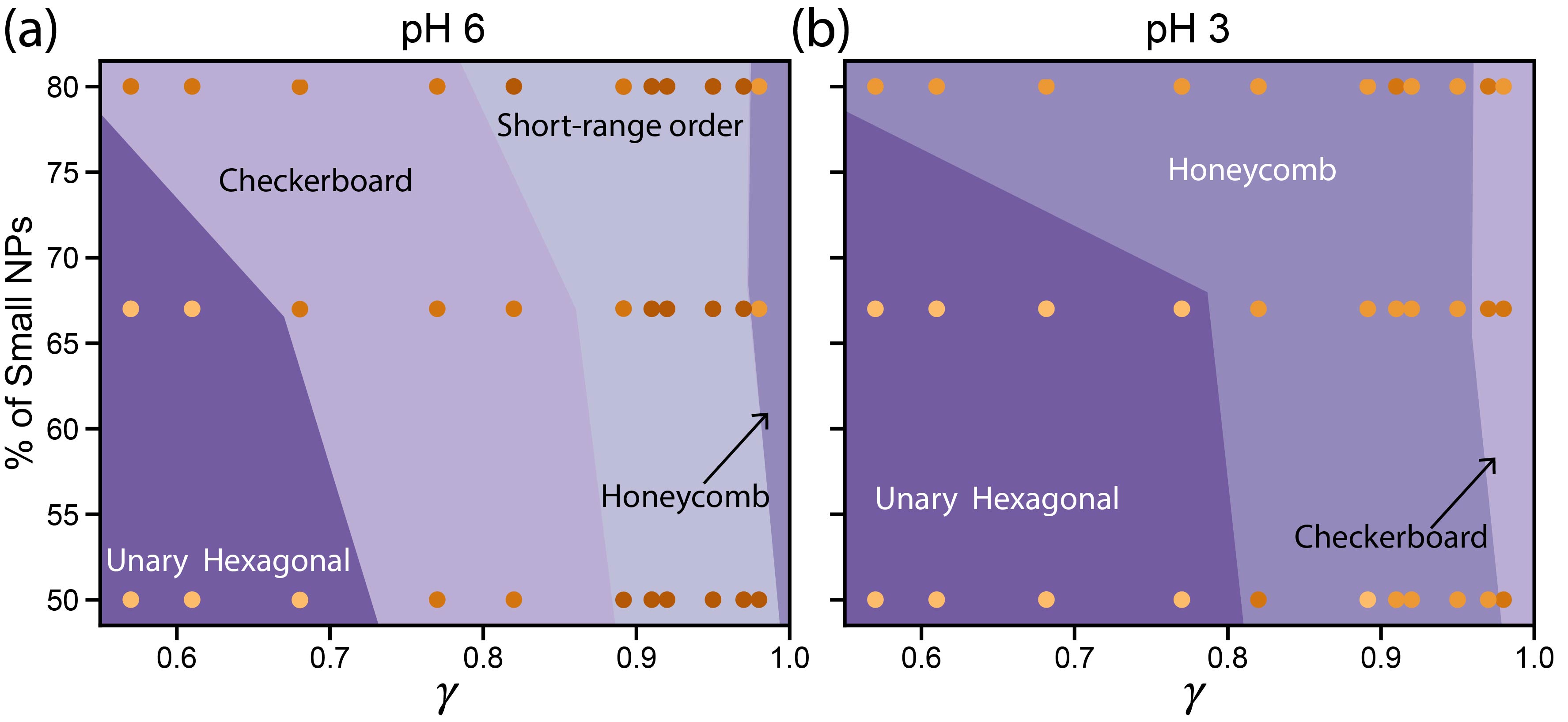}
    \caption{\textbf{Phase Behavior of Binary PEG‐AuNP Superlattices.} Preliminary phase diagrams illustrating the assembled lattice structures as a function of $\gamma$ (the size/diameter ratio of the two grafted AuNP species) versus their nominal mixing ratio. The phase diagrams are presented for two distinct pH values: 6 (left panel) and 3 (right panel). At small $\gamma$, both pH values favor the crystallization of a single-component phase at the surface, exhibiting a simple hexagonal lattice. As $\gamma$ increases, a checkerboard-square lattice becomes stable at pH 6, while a honeycomb structure is prevalent at pH 3. At the highest $\gamma$ values explored, the checkerboard lattice dominates at pH 3, whereas at pH 6, short-range order is generally observed, with the honeycomb structure reappearing only within a narrow range of $\gamma$ close to unity. These diagrams suggest a form of structural reversal between the honeycomb and checkerboard lattices as a function of pH and $\gamma$. The phase diagram is constructed using data from Figs. S3-S11}
    \vspace{-0.4 cm}
 	\label{fig:phase_diagram} 
 \end{figure*}
 
To overcome such kinetic competition and achieve a binary superlattice at the interface, we increased the concentration of the less surface-active component (the shorter-chain grafted AuNPs) to enable it to effectively compete for interfacial space. In fact, as shown in Figure~\ref{fig:checkerboard_hexagonal_unary} (d-e), the same binary mixture, when prepared at a nominal mixing ratio of 4:1, orders into a checkerboard superstructure, leveraging charge balance at pH 6 and pH 4. However, upon further lowering the pH to 3 (Fig.~\ref{fig:checkerboard_hexagonal_unary} (f)), the GISAXS pattern reveals a mixed structural state, indicating the coexistence of both the checkerboard lattice and the original unitary hexagonal superstructure.
 
To further explore the effect of polymer MW disparity, we examined a binary mixture of COOH‐PEG2k-Au5 and NH$_2$‐PEG10k-Au10 at a 4:1 nominal ratio (additional ratios in the Supporting Information). Fig. \ref{fig:Checkerboard_hexagonal} shows GISAXS diffraction line-cuts recorded at pH 6, 4, and 3. At pH 6 (Fig.~\ref{fig:Checkerboard_hexagonal}(a)), the pattern corresponds to a well-defined checkerboard superlattice, with the strongest Bragg peak being resolution-limited, indicating long-range correlations and an average crystal size of $\gtrsim 600$ nm. Lowering the pH to 4 (Fig. \ref{fig:Checkerboard_hexagonal} (b)) causes a slight weakening of this order. Upon further acidification to pH 3 (Fig. \ref{fig:Checkerboard_hexagonal}(c)), the structure reorganizes into a body‐centered honeycomb lattice with a quasi‐stoichiometric A$_2$B composition. This transition occurs because protonation of the –NH\textsubscript{2} termini at a lower pH increases their solubility, reducing their interfacial occupancy and allowing a higher relative density of COOH-PEG2k-Au5 to drive the formation of the A$_2$B hexagonal superstructure. We note that checkerboard and hexagonal lattices are symmetry-incompatible; therefore, the transition is not a continuous one. Thus, after each pH change, we briefly stir to erase structural memory (as described in detail in the Methods section) and allow the interface to re-equilibrate to the thermodynamically selected lattice consistent with the particle charge states.

Extensive data on the various mixtures are provided in the SI, including all those discussed in the main text. The corresponding diffraction line-cuts and fitted profiles are reproduced in the SI, along with a comprehensive list of refined parameters. These parameters include lattice constants and nearest-neighbor distances for each condition and structure. Detailed information on the structure factors employed in the fitting procedure is also presented in the SI.
\vspace{-0.5 cm}

\section{Conclusions}
\vspace{-0.2 cm}
We have successfully demonstrated control over the 2D assembly and ordering of binary AuNP systems at the vapor-liquid interface by functionalizing them with water-soluble PEG bearing either carboxylic acid (-COOH) or amine (-NH$_2$) end groups. The electrostatic interactions between these charged end groups, which can be effectively tuned by varying the pH of the mixed suspension, serve as a key driving force for the self-assembly process. Furthermore, the MW of the grafted PEG (2, 5, and 10 kDa), which influences the hydro-dynamic size of the particles, and the core size of the AuNPs (5 and 10 nm) provide additional parameters for controlling the resulting superlattice structures. We have achieved a diverse range of ordered 2D lattices, including checkerboard, simple square, and body-centered honeycomb configurations. Our findings reveal that manipulating the MW of grafted PEG, in both size-mismatched and size-matched AuNP systems, drives the formation of robust superlattices. The effect of MW on the conformation of the polymer and the grafting density underscores the pivotal role of these parameters in the direction of assembly. Moreover, we observed pH-induced structural transitions between these different lattice types for specific combinations of NP size and polymer length, highlighting a versatile approach for the programmable assembly of colloidal crystals with tailored architectures. Finally, we have established preliminary phase diagrams that illustrate the relationship between the formed structure as a function of $\gamma$ (the diameter ratio of the two AuNP species) and their mixing ratio in suspension, as shown in Fig. \ref{fig:phase_diagram} (the map is based on a limited sampling of $\gamma$ and mixing ratios). At small $\gamma$, both pH values favor the crystallization of a single-component phase on the surface, exhibiting a simple hexagonal lattice. As $\gamma$ increases, a checkerboard-square lattice becomes stable at pH 6, while a honeycomb structure is prevalent at pH 3. At the highest $\gamma$ values explored, the checkerboard lattice dominates at pH 3, whereas at pH 6, short-range order is generally observed, with the honeycomb structure reappearing only within a narrow range of $\gamma$ close to unity.
\vspace{-0.6 cm}

\section{Materials and Methods}
\vspace{-0.3 cm}
\subsection{Materials}
\vspace{-0.1 cm}
Citrate-stabilized gold nanoparticles (AuNPs) with nominal core diameters of 5 and 10 nm were obtained from Ted Pella Inc., and their size distributions were independently verified by transmission electron microscopy (TEM) (shown in Figs. S1-S2) and small-angle X-ray scattering (SAXS).\cite{nayak2023tuning} Thiol-terminated polyethylene glycols (HS-PEG-X, where X = COOH or NH\textsubscript{2}), with average molecular weights of 2, 5, or 10 kDa, were purchased from Creative PEGworks (NC, USA) and used as received. Hydrochloric acid (HCl) was sourced from Fisher Scientific without further purification. All experiments employed Milli-Q water (18.2 M$\Omega$·cm at 25 \textdegree{}C).
\vspace{-0.5cm}
\subsection{Ligand Exchange and PEG Grafting}
\vspace{-0.1 cm}
AuNPs were functionalized with HS-PEG-COOH or HS-PEG-NH\textsubscript{2} via a standard ligand-exchange protocol.\cite{zhang2017macroscopic,nayak2023assembling} Briefly, PEG ligands were dissolved in Milli-Q water and added in molar excess to the AuNP suspensions at particle-to-PEG ratios of 1:1500 (5 nm cores) or 1:6000 (10 nm cores). The mixtures were rotated ($\sim$35 RPM) overnight on a Roto-Shake Genie (Scientific Industries, NY, USA). Excess PEG was removed by three successive centrifugation steps (5 nm: 21,000 ×g, 90 min; 10 nm: 20,000 ×g, 75 min), and the pellets were redispersed in Milli-Q water to generate stock suspensions. Nanoparticle concentrations were determined by UV–vis absorbance (NanoDrop One, Thermo Fisher Scientific) and adjusted to ~80 nM (5 nm) or ~20 nM (10 nm) as previously described.\cite{nayak2023tuning}

In the nomenclature used here, “PEG-AuNP” refers to any PEG-grafted nanoparticle, whereas the label \textit{x}-PEG\textit{y}-Au\textit{z} specifies:  
- \textit{z} = core diameter (nm),  
- \textit{y} = PEG molecular weight (kDa),  
- \textit{x} = terminal group (COOH or NH\textsubscript{2}).  
For example, COOH-PEG5k-Au10 denotes a 10 nm AuNP functionalized with 5 kDa PEG terminating in –COOH.

For clarity, the schematics accompanying the figures depict the nanoparticles as idealized hard spheres to emphasize the high crystalline order. This representation is justified by SAXS measurements of the dispersed particles, which are well described by a spherical form factor, and by TEM images confirming high sphericity with negligible shape anisotropy.

\vspace{-0.4cm}
\subsection{Dynamic Light Scattering and Zeta Potential}
\vspace{-0.1 cm}
Hydrodynamic diameters (D\textsubscript{H}) and zeta potentials ($\zeta$) of the PEG–AuNPs were measured using a Malvern Zetasizer Nano ZS90 with Zetasizer software (Malvern, UK). Successful grafting of COOH- and NH$_2$-terminated PEG chains was confirmed by systematic increases in D\textsubscript{H} relative to the ungrafted cores and by the emergence of oppositely signed surface charges. Table~S1 summarizes the measured D\textsubscript{H} and $\zeta$ values for all PEG-grafted AuNPs used in this work together with the nomenclature adopted throughout the manuscript. These measurements were used to compute the size ratio, $\gamma$, reported alongside the phase diagram in Fig.~\ref{fig:phase_diagram}. Further experimental details, data processing procedures, and representative DLS traces are provided in a recently published manuscript Ref. [\!\!\citenum{nayak2025valencefree}].

\vspace{-0.5cm}
\subsection{Sample Preparation for Liquid-Surface X-Ray Diffraction}
\vspace{-0.1 cm}
For each experiment, PEG–AuNPs suspensions were combined at the desired molar ratio to a final volume of 2 mL and equilibrated for \mbox{$\sim$15 min} before transfer to a stainless–steel trough housed in a water–saturated, helium–purged chamber on the liquid–surface spectrometer.\cite{pershan2012liquid,Vaknin2003a} 
To achieve successive target pH values with minimal change in suspension volume and nanoparticle concentration, concentrated HCl stock solutions (10 mM, 100 mM, and 1 M) were used for incremental titration. A calculated volume of HCl was introduced as small aliquots (\mbox{$\sim$20 $\mu$L} per 2 mL AuNP suspension) at a slow, controlled rate (\mbox{$\sim$5 $\mu$L} per drop) using a calibrated micropipette. The dispensing tip was positioned at multiple points along the liquid surface to promote uniform distribution while avoiding local overacidification. After each addition, the suspension was gently mixed by pipetting for \mbox{$\sim$30 s} to ensure homogeneity without introducing air bubbles or bulk turbulence. The sample was then allowed to equilibrate for \mbox{$\sim$20 min} prior to GISAXS/XRR measurements. This titration–equilibrate–measure cycle was repeated within the same suspension to construct the pH series. Specifically, for a 2 mL suspension, \mbox{20 $\mu$L} of 10 mM HCl was added to reach \mbox{$[\mathrm{HCl}]\!\approx\!0.1$ mM} (pH $\sim$ 4); \mbox{18 $\mu$L} of 100 mM HCl was added to reach \mbox{$[\mathrm{HCl}]\!\approx\!1$ mM} (pH $\sim$ 3); and \mbox{19 $\mu$L} of 1 M HCl was added to reach \mbox{$[\mathrm{HCl}]\!\approx\!10$ mM} (pH $\sim$ 2) within the same suspension.

\vspace{-0.4cm}
\subsection{In Situ Liquid-Surface X-Ray Reflectivity and GISAXS}
\vspace{-0.1 cm}
Synchrotron-based in situ measurements were carried out at the Open Platform Liquid Surfaces (OPLS) end station of the Soft Matter Interfaces beamline (SMI), NSLS-II, Brookhaven National Laboratory, using 14.4 keV ($\lambda =$ 0.0861 nm) incident radiation. Specular X-ray reflectivity (XRR) profiles, \(R(Q_z)\), were recorded at grazing incidence and normalized by the theoretical Fresnel reflectivity, \(R_F\). Electron density profiles, \(\rho(z)\), were extracted by fitting \(R/R_F\) data using Paratt’s recursive algorithm.\cite{Nielsen2011} Concurrently, grazing-incidence small-angle X-ray scattering (GISAXS) patterns were collected on an area detector to probe the in-plane ordering of the nanoparticle monolayer. Two-dimensional intensity maps (\(Q_{xy}\) vs.\ \(Q_z\)) were processed by integrating line cuts over \(Q_z = 0.02\text{--}0.04\)\,\AA\(^{-1}\) to yield \(I(Q_{xy})\) (see SI Fig. S18); diffraction peaks were indexed by Miller indices \((h,k)\) corresponding to the observed 2D lattice symmetry.\cite{kim2022binary,kim2022lamellar,zhang2017macroscopic}GISAXS line-cut profiles were fit with sums of Lorentzian line shapes of calculated intensities using the structure factor (with a constant background) to extract peak positions, full widths at half maximum (FWHM), and amplitudes. These parameters were then used to construct an idealized structure factor for each assigned lattice. Full details of the fitting protocol, parameter constraints, and the calculation of the idealized structure factor are provided in the SI.

\vspace{-0.4cm}
\section{Supporting Information}\label{SI}
\vspace{-0.3cm}
Supporting Information: 
The Supporting Information is available free of charge on the Publisher's website at\\ 
DOI: doi.org/10.1016/j.mtnano.2025.100734

Supplementary data: DLS \& $\zeta$-potential, TEM imaging, GISAXS structure factor, additional GISAXS \& XRR data, and ED profiles.

\vspace{-0.5cm}
\section{Data Availability}
\vspace{-0.3cm}
The findings of this study are supported by data found within the article and the provided Supplementary Information. Further relevant information and source data can be obtained from the link below. DOI: doi.org/10.7910/DVN/WXFOYO

\vspace{-0.5cm}
\section{Acknowledgements}
\vspace{-0.3cm}
This work was supported by the U.S. Department of Energy (DOE), Office of Science, Basic Energy Sciences, Materials Science and Engineering Division. The research was performed at the Ames National Laboratory, which is operated for the U.S. DOE by Iowa State University under contract No. DE-AC02-07CH11358. Part of this research used the Open Platform Liquid Surfaces (OPLS) end station of the Soft Matter Interfaces Beamline (SMI, Beamline 12-ID) of the National Synchrotron Light Source II, a U.S. DOE Office of Science User Facility operated for the DOE Office of Science by Brookhaven National Laboratory under Contract No. DE-SC0012704.

\normalem
\vspace{-0.5cm}
\bibliography{Ref.bib}

\end{document}